\documentclass[10pt,superscriptaddress,twocolumn,amsmath,amssymb,aps,prl,showpacs]{revtex4}
\usepackage{amsmath,amssymb,mathrsfs}
\usepackage[dvips]{graphicx}
\usepackage{hyperref}
\usepackage{paralist}

\def\be{\begin{equation}}
\def\ee{\end{equation}}
\def\bea{\begin{eqnarray}}
\def\eea{\end{eqnarray}}

\usepackage{stmaryrd}
\usepackage{amssymb}
\usepackage{amsmath}
\usepackage{amsfonts}
\usepackage{amsthm}
\usepackage{color}
\usepackage{epsfig}
\usepackage{bm}
\usepackage{rotating}
\usepackage{mathrsfs}
\usepackage{concmath}
\usepackage[T1]{fontenc}
\usepackage{cleveref}
\usepackage{microtype}
\usepackage{subfigure}

\begin{document}

\title{A unified approach to the thermodynamics and quantum scaling functions of  one-dimensional  strongly attractive $SU(w)$ Fermi  Gases}

\author{Yi-Cong Yu}
\affiliation{State Key Laboratory of Magnetic Resonance and Atomic and Molecular Physics,
Wuhan Institute of Physics and Mathematics, Chinese Academy of Sciences, Wuhan 430071, China}
\affiliation{University of Chinese Academy of Sciences, Beijing 100049, China.}

\author{Xi-Wen Guan}
\email[]{xiwen.guan@anu.edu.au;xwe105@wipm.ac.cn}
\affiliation{State Key Laboratory of Magnetic Resonance and Atomic and Molecular Physics,
Wuhan Institute of Physics and Mathematics, Chinese Academy of Sciences, Wuhan 430071, China}
\affiliation{Center for Cold Atom Physics, Chinese Academy of Sciences, Wuhan 430071, China}
\affiliation{Department of Theoretical Physics, Research School of Physics and Engineering,
Australian National University, Canberra ACT 0200, Australia}

\pacs{05.30.Fk, 02.30.Ik,03.75.Ss}

\date{\today}

\begin{abstract}

In this letter we present a unified derivation of the pressure  equation of states, thermodynamics and scaling functions for the one-dimensional (1D) strongly attractive  Fermi gases with $SU(w)$ symmetry. These  physical quantities provide a rigorous understanding on a universality class of quantum criticality characterised by the critical exponents $z=2$ and correlation length exponent $\nu=1/2$. Such a universality class of quantum criticality  can occur when  the Fermi sea of one branch of charge bound states starts to fill or become  gapped  at zero temperature.  The quantum critical cone   can be determined through  the double peaks     in  specific heat  which  serve to mark two crossover temperatures fanning out from the critical point.  Our method opens to further  study on  quantum phases and  phase transitions in strongly  interacting fermions with large $SU(w)$  and non-$SU(w)$ symmetries in one dimension. 

\end{abstract}

\maketitle

The experimental realization of the 1D quantum gases, such as repulsive Bose gases \cite{Exp-B1,Exp-B2,Exp-B3,Exp-B4,Exp-B5,Exp-B6}, Yang-Gaudin model \cite{Liao:2010,Wenz:2013}, multicomponent attractive Fermi gases \cite{Pag14},  has provided a remarkable test ground for exactly solvable models. The mathematical theory of Bethe ansatz integrable models has become testable in ultracold atoms. The Bethe anssatz  has also found  success for other significant models like the Kondo physics \cite{Andrei:1983}, BCS pairing model \cite{Dukelsky:2004}, strongly correlated electronic systems \cite{Ess05,Takahashi,WangYP:2015}, spin ladders \cite{Wang99,Batchelor:2007} and quantum degenerate  gases \cite{Cazalilla,GuaBL13}. 

Recent studies of the 1D Fermi gases with high spin symmetries \cite{GuaBL13,Yu:2016,WuHZ03,CazR14,Schlottmann:1993,LeeGBY11,Jiang:2016,He:2010,Oelkers:2006} has given  many theoretical predictions on the existence of  bound states of multiparticles, quantum liquids and phase transitions. In this regard, exploring exactly solvable models of interacting fermions with high mathematical symmetries is highly desirable in order to understand new phases of matter.  However,  the thermodynamic properties of exactly solvable models with high symmetries at finite temperatures are notoriously difficult to extract and  present a formidable challenge. 
Building on Yang-Yang thermodynamic Bethe ansatz equations,  here we present a unified approach to the thermodynamics and quantum critical scalings in  1D strongly attractive  Fermi gases with $SU(w)$ symmetry.  Analytical results of the equation of states(EOS), dimensionless ratios and  scaling functions of thermal and magnetic properties  provide a rigorous understanding on a universality class of quantum criticality of free fermions. The quantum critical region   can be determined through  the double maxima    in  specific heat  which  characterize the  two crossover temperatures fanning out from the critical point.  These results suggest to experimentally test the universal nature of multicomponent quantum liquids  through the  1D strongly attractive  Fermi gases of ultracold atoms  \cite{Pag14}.


{\bf The model.} The 1D  $SU(w)$ Fermi gases  with $\delta$-function interaction
confined to length $L$  is described by the   following Hamiltonian \cite{Yang,Gaudin,Sut68,Tak70}
\begin{eqnarray}
 {H}&=&-\frac{\hbar^{2}}{2m}\sum_{i=1}^{N}\frac{\partial^{2}}{\partial
x_{i}^{2}}+g_\mathrm{1D}\sum_{1\leq i<j\leq
N}\delta(x_{i}-x_{j})\nonumber\\
&&- E_z- \mu N, \label{Ham}
\end{eqnarray}
and with the chemical potential $\mu$ and the  effective Zeeman energy $E_{z} =
\sum_{r=1}^{w}\frac{1}{2}r(w-r) n_{r}H_{ r}$.
Here  $N$ is the total particle number.
There are $w$ possible hyperfine states $|1\rangle, |2\rangle, \ldots,
|w\rangle$ that the fermions can occupy.
Experimentally, $g_\mathrm{1D}=-2\hbar^2/m a_\mathrm{1D}$, with $a_\mathrm{1D}$ the effective scattering length in 1D \cite{Ols98},  can be tuned from a weak interaction to a strong coupling regime  via Feshbach resonances.
For our convenience, from now on, we choose our units such that $\hbar^2=2m=1$ unless we particularly use the units.  
In this model,  the two-body charge bound states involve the Bethe ansatz roots
$\{\lambda_j\pm i c/2\}, j=1...M_2$ and the three-body bound states
$\{\lambda_j \pm i c, \lambda_j\},j=1...M_3$, and so on, where $M_2$ and $M_3$
are the numbers of charge bound states and three-body bound states,
respectively \cite{LeeGBY11}.  The thermodynamics of the model  are determined by the effective external fields $H_r$,
chemical potential, interaction between different particles  and spin
wave fluctuations.

{\bf The TBA  equations.} The thermodynamics of  the Hamiltonian (\ref{Ham})  is determined by  the following TBA equations \cite{Takahashi,Tak70,LeeGBY11}
\begin{small}
\begin{eqnarray}
\epsilon^{(r)}(k)&=&rk^2 - r{\mu}- H_r - \frac{r(r^2-1)c^2}{12}\nonumber \\
&&
-\sum_{q=1}^w \hat{a}_{rq}  \ast F[\epsilon^{(q)}]
+\sum_{q=1}^{\infty} a_q \ast F[\eta_{r,q}],\label{TBA}
\end{eqnarray}
\begin{eqnarray}
\eta_{r,l}(k)& =& l \cdot \left(2 H_r - H_{r - 1} - H_{r + 1} \right) 
- a_l \ast F[\epsilon^{(r)}] \nonumber \\
&& - \sum_{q=1}^{\infty} U_{lq} \ast F[\eta_{r,q}] 
 + \sum_{q=1}^{\infty} S_{lq} \ast F[\eta_{r-1,q}]\nonumber \\
&&+ \sum_{q=1}^{\infty} S_{lq} \ast F[\eta_{r+1,q}],\label{TBA-spin}
\end{eqnarray}
\end{small}
where we denote 
\begin{eqnarray}
a_n(x)&=&\frac{1}{2\pi}\frac{n|c|}{(nc/2)^2+x^2},\nonumber  \\
\hat{a}_{lj}(x)&=&\sum_{q=1,2q \neq l+j}^{min{(l,j)}} a_{l+j-2q}(x),\label{definition_F}\nonumber\\
F[\varepsilon] &\triangleq & -T \ln[1+\exp(-\frac{\varepsilon}{T})].\nonumber
\end{eqnarray}
In the above equations, $\ast$ denotes the convolution 
$(f \ast g)(\lambda) = \int_{-\infty}^\infty f(\lambda-\lambda') g(\lambda') d\lambda'$ and 
the functions   $U_{lj}(x)$ and $S_{lj}(x)$ are given in \cite{LeeGBY11}.
From the dressed energies $\epsilon^{(r)}(k)$ for bound states of $r$-atoms with $r=1,\cdots, w$, one can obtain the pressure 
\begin{align}
p=\sum_{r=1}^w \frac{rT}{2\pi} \int_{-\infty}^{\infty} 
  dk \ln(1+e^{-\epsilon^{(r)}/T}).
  \label{eq:pressure}
\end{align}
The summation of the pressures of all charge bound states services as the EOS,
from which we can obtain  full thermodynamics of the model at the  temperatures ranging from zero to high.  This form of the EOS gives rise to the additivity nature of quantum liquids in low temperatures \cite{Yu:2016}.

\begin{figure}[t]
\begin{center}
  \includegraphics [width=1.0\linewidth]{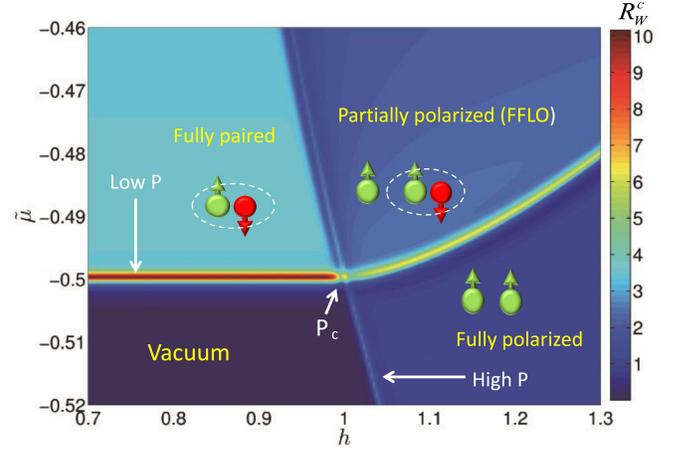}
\end{center}
 \caption{\footnotesize{Contour plot of the Wilson ratio $R_{\rm W} ^{\rm c}$ in $\tilde{\mu}-h$ plane at   temperature $T=0.001\varepsilon_2/k_B$. The calculation of the Wilson ratio (\ref{ratio2}) was carried out by numerically solving the TBA equations (\ref{TBA}).
 Here $\tilde{\mu}=\frac{\mu}{\frac{1}{2}|c|^2}$ and $h=\frac{H}{\frac{1}{2}|c|^2}$ are  the dimensionless chemical potential and magnetic field.
$P_c $ stands for the critical polarization. The $P=(N_{\uparrow}-N_{\downarrow})/(N_{\uparrow}+N_{\downarrow})$ denotes the particle number imbalance.
The Wilson ratio gives different constant values which characterize three  Luttinger liquid  phases of fully-paired state, partially-polarized Fulde-Ferrell-Larkin-Ovchinnikov (FFLO) like state and the fully-polarized normal Fermi gas.
The Wilson ratio remarkably  displays two distinct  plateaus of the integers 1 and 4 in strong coupling limit. }
  }
 \label{fig:Phase_diagram}
\end{figure}

We are interested in the low temperature behaviour of interacting fermions with high symmetries  in 1D. 
We can see from  the TBA equations (\ref{TBA-spin}) that  the  fermomagnetic ordering (the second term in (\ref{TBA-spin})) drives the spin contributions $\eta_{r,l}(k)$ to the dressed energies of charge bound states exponentially small in strongly attractive regimes.  Consequently, we can ignore all the string contributions in 
the above TBA equations when temperature is much less than the binding energies of the charge bound states $\varepsilon_r=\frac{1}{48} r \left( r^2 -1\right) g_\mathrm{1D}^2$. In the recent study \cite{Yu:2016}, it has been proved  that the dimensionless Wislon  ratios, i.e. either the ratio of the susceptibility $\chi$   to the specific heat $c_V$ divided by the temperature
\begin{equation}
R_{\rm W}^{\rm s}=\frac{4}{3}\left(\frac{\pi k_B}{\mu_B g }\right)^2\frac{\chi}{c_V/T},
\label{ratio}
\end{equation}
or the  ratio  of  the compressibility $\kappa$  to the specific heat $c_V$ divided by the temperature
\begin{equation}
  R_{\rm W}^{\rm c}=\frac{k_B^2\pi^2 }{3}\frac{\kappa}{c_V/T},
\label{ratio2}
\end{equation}
essentially captures the  quasiparticle   nature of Fermi liquid  \cite{Sommerfeld,Wilson1975},  as its value characterizes the interacting effect in the Fermi liquid.
Here $k_B$ is Boltzmann's constant, $\mu_B$ is the Bohr magneton and $g$ is the Lande factor.
The two types of dimensionless ratios (\ref{ratio}) and  (\ref{ratio2})   characterise   a competition between the   fluctuations of two thermodynamic quantities.
Thus  a constant  Wilson ratio    implies that  the two types of  fluctuations are on an equal footing with respect to the temperature, regardless of the microscopic details of many-body systems \cite{Wang98,Guan-PRL,Yu:2016}. In Fig.~\ref{fig:Phase_diagram}, we demonstrate the compressibility Wilson ratio elegantly maps out the full  phase digram of the $SU(2)$ Fermi gas \cite{Yang,Gaudin} with a strong attraction at $T=0.001\varepsilon_2/k_B$, where $\varepsilon_2$ is the binding energy of a  bound pair. This  turns out that the low temperature $SU(2)$ TBA equations (\ref{TBA}) provide rigorous results of quantum liquid behaviour and quantum criticality,  prompting  us  to explore universal thermodynamics and quantum scaling functions for the high symmetry Fermi gases with strong attractions through the $SU(w)$ TBA equations.

Without loss of accuracy at  low temperature thermodynamics, we  simplify the TBA equations (\ref{TBA})  into  the following form 
\begin{align}
\epsilon^{(r)}(k)=V_r-\sum_{q=1}^w \hat{a}_{rq} \ast F[\epsilon^{(q)}] , 
\qquad r=1,2,\cdots,w
\label{eq:TBA}
\end{align}
with $
V_r=rk^2-r\mu-H_r-\frac{1}{12}r(r^2-1) c^2$,
where  the last effective field $H_w$ is set to zero due to the spin singlet charge bound states. 
For example,
 the $SU(3)$ case, i.e. $w=3$ in the Eq. (\ref{eq:TBA})
 determines  the low temperature properties of the 1D  three-component strong interacting fermions \cite{Guan:2008,Foerster:2012}. 

{\bf Equation of states.}
In the strong coupling region, i.e. $|c| \gg E_F $ ($E_F$ is the Fermi energy) and at  low temperatures, we have the following expansion form
\begin{eqnarray}
&&-a_n*F[\epsilon^{(m)}](k) \approx \frac{n|c|}{(nc/2)^2+k^2}\frac{p^{(m)}}{m}\nonumber\\
          &&    +\frac{1}{2\pi}\frac{32}{3m\sqrt{m}}\frac{1}{n^3|c|^3}
              \Gamma(\frac{5}{2})T^{\frac{5}{2}}{\rm Li}_{\frac{5}{2}}(-e^{A^{(m)}/T}),
              \label{eq:important}
\end{eqnarray}
which is obtained  by integration by parts. In (\ref{eq:important}) the 
$\rm{Li(x)}$ is the polylog function. Substituting  this relation into 
the TBA equations \cref{eq:TBA}, we have 
\begin{eqnarray}
\epsilon^{(r)}(k)&=&V_r+\sum_{m=1}^w \sum_{\mbox{\tiny$\begin{array}{c}
q=1\\
2q < r+m\end{array}$}}^{min(r,m)} \frac{4}{|c|}
\frac{p^{(m)}}{m(r+m-2q)} \notag \\
&&-\sum_{m=1}^w \sum_{\mbox{\tiny$\begin{array}{c}
q=1\\
2q < r+m\end{array}$}}^{min(r,m)} \frac{16}{|c|^3}
\frac{p^{(m)}}{m(r+m-2q)^3}k^2\nonumber \\
&&+\sum_m^w Z_{rm}\frac{T^{5/2}}{|c|^3} {\rm Li}_{\frac{5}{2}} (-e^{A^{(m)}/T}).
\label{eq:expand_dressed_energy}
\end{eqnarray}
In the above equation, we defined matrix,
\begin{align}
Z_{rm}=\sum_{\mbox{\tiny$\begin{array}{c}
q=1\\
2q < r+m\end{array}$}}^{min(r,m)}  \frac{4}{\sqrt{\pi}} \frac{1}{(r+m-2q)^3 m^{3/2}}
\end{align}
and the $A^{(m)}$ collects  the constant  terms (terms are independent of the $k$) in the dressed energy $\epsilon^{(m)}(k)$. We would like to mention that the polylog functions involve different modes of generating functions of Fermi integrals. In contrast to the Somerfield expansion with respect to the powers of the temperature $t$, here the polylog functions contain enough thermal and quantum fluctuations   that are acquired by the quantum criticality. Therefore, the polylog functions essentially characterise the singular behaviour of the 1D strongly interacting fermions even near quantum phase transitions. 
If we  consider the first three orders in  the pressure, only the constant term and the 
quadratic term ($O(k^2)$ terms) contribute to the thermodynamic quantities, and we can safely 
drop  high order terms  in $k$ .
Explicitly, we express the dressed energy as 
\begin{align}
\epsilon^{(r)}=r D_r k^2- A^{(r)},
\end{align}
Integrating  eq.(\ref{eq:pressure}) by parts we have
\begin{equation}
p^{(m)}=-\frac{\sqrt{m}}{2\sqrt{\pi D_m}} T^{\frac{3}{2}} {\rm Li}_{\frac{3}{2}}
    (-e^{A^{(m)}/T}).  \label{eq:pressure_result}
\end{equation}
The parameter $D_r$ is a  modification to  the quadratic term in  the dressed energy $\epsilon^{(m)}(k)  $ and could be read off  from the Eq.(\ref{eq:expand_dressed_energy})
\begin{align}
D_r=1-\sum_{m=1}^w \sum_{\mbox{\tiny$\begin{array}{c}
q=1\\
2q < r+m\end{array}$}}^{min(r,m)} \frac{16}{|c|^3}
\frac{p^{(m)}}{rm(r+m-2q)^3} .
\end{align}
Here we only consider the first three orders in the dressed energy equations, thus  $D_r \approx 1$, and the EOS becomes
\begin{eqnarray}
p^{(r)}&=&-\frac{\sqrt{r}}{2\sqrt{\pi}} T^{\frac{3}{2}} {\rm Li}_{\frac{3}{2}}
    (-e^{\frac{A^{(r)}}{T}}), \label{eq:pressure2} \\
A^{(r)}&=& A_0^{(r)}-\sum_{m=1}^N \frac{D_{rm} p^{(m)}}{|c|} 
- \frac{T^{\frac{5}{2} }Z_{rm} {\rm Li}_{\frac{5}{2}}
\left(-e^{\frac{A^{(m)}}{T}} \right)}{|c|^3}, \nonumber
\end{eqnarray}
where $A^{(r)}_0$ and $D_{rm}$ are determined by the Eq.(\ref{eq:expand_dressed_energy})
\begin{eqnarray}
A^{(r)}_0&=&r \mu+H_r+\frac{1}{12}r(r^2-1)c^2,\nonumber \\
D_{rm}&=& \sum_{\mbox{\tiny$\begin{array}{c}
q=1\\
2q < r+m\end{array}$}}^{min(r,m)} \frac{4}
{m(r+m-2q)}. \nonumber
\end{eqnarray}
In order to simplify the EOS,  we define the dimensionless quantities and  parameters
\begin{eqnarray}
\tilde{p}^{(r)}&=&\frac{p^{(r)}}{|c|^3},\,\,
\tilde{A}^{(r)}=\frac{A^{(r)}}{|c|^2},\nonumber\\
\tilde{\mu}&=&\frac{\mu}{|c|^2},\,\, 
{h_r}=\frac{H_r}{|c|^2},\,\,
t=\frac{T}{|c|^2}.
\label{eq:su3AP_renorm}
\end{eqnarray}
Then the dimensionless EOS is given by 
\begin{align}
\tilde{p}^{(r)}=&-\frac{\sqrt{r}}{2\sqrt{\pi}} t^{\frac{3}{2}} {\rm Li}_{\frac{3}{2}}
    (-e^{\frac{\tilde{A}^{(r)}}{t}}), \label{eq:su3pressure3} \\
\tilde{A}^{(r)}=& \tilde{A}_0^{(r)}-\sum_{m=1}^N D_{rm} \tilde{p}^{(m)} 
-Z_{rm} t^{\frac{5}{2}} {\rm Li}_{\frac{5}{2}}
(-e^{\frac{\tilde{A}^{(m)}}{t}}).\nonumber
\end{align}
where  $\tilde{A}_0^{(r)}=r \tilde{\mu}+h_r+\frac{1}{12}r(r^2-1)$. 
For simplifying our notations, we further define matrices 
\begin{eqnarray}
&&(\bm{{\rm\ Li}_{s}})_{rm}={\rm Li}_{s}
    (-e^{\tilde{A}^{(r)}/t})\delta_{rm},\,\,
    (\bm{D})_{rm}=D_{rm},\nonumber\\
 &&     (\hat{\bm{Z}})_{rm}=Z_{rm},   (\bm{\tilde{A}})_{r1}=\tilde{A}^{(r)},\,\,
 (\bm{\tilde{p}})_{r1}=\tilde{p}^{(r)},\nonumber\\
&&  (\bm{M_r})_{rm}=\frac{\sqrt{r}}{2\sqrt{\pi}}\delta_{rm},  (\bm{F_s})_{rm}\triangleq t^s {\rm Li}_s(-e^{\tilde{A}^{(r)}/t})\delta_{rm},\nonumber\\
&&(\bm{f_s})_{rm}\triangleq t^s {\rm Li}_s(-e^{\tilde{A}_0^{(r)}/t})\delta_{rm}.
\end{eqnarray}
$\bm{{\rm Li}_{\frac{5}{2}}}$, $\bm{{\rm Li}_{\frac{3}{2}}}$,
     $\bm{M_r}$,$\bm{D}$ and $\bm{\tilde{Z}}$ are square matrices ( or the  column matrices  when they are at the most right of the related terms),  $\bm{\tilde{p}}$ and 
     $\bm{\tilde{A}}$ are column matrices.
   For example, in the $SU(2)$ case
\begin{align}
\bm{D} =
\left(
\begin{array}{cc}
 0 & 2 \\
 4 & 1 \\
\end{array}
\right),
\qquad
\bm{Z}=
\left(
\begin{array}{cc}
 0 & \frac{\sqrt{2}}{\sqrt{\pi}} \\
 \frac{4}{\sqrt{\pi}} & \frac{1}{4\sqrt{2\pi}} \\
\end{array}
\right),\nonumber
\end{align}
and for $SU(3)$ case,
\begin{align}
\bm{D} =
\left(
\begin{array}{ccc}
 0 & 2 & \frac{2}{3} \\
 4 & 1 & \frac{16}{9} \\
 2 & \frac{8}{3} &  1 \\
\end{array}
\right),\,\,
\bm{Z}=
\left(
\begin{array}{ccc}
 0 & \frac{\sqrt{2}}{\sqrt{\pi}} & \frac{1}{6\sqrt{3\pi}}\\
\frac{4}{\sqrt{\pi}} & \frac{1}{4\sqrt{2\pi}} & \frac{112}{81\sqrt{3\pi}} \\
\frac{1}{2\sqrt{\pi}} &\frac{56}{27\sqrt{2\pi}} &\frac{\sqrt{3}}{16\sqrt{\pi}}\\
\end{array}
\right).\nonumber
\end{align}
With the help of these notations,
we rewrite  a unified expression of EOS for the $SU(w)$ strongly attractive Fermi gases
\begin{align}
\tilde{\bm{p}}
=&-\bm{M_r}t^{3/2} \bm{{\rm Li}_{\frac{3}{2}}}
\label{eq:result_p},  \\
\bm{\tilde{A}}
=& \bm{\tilde{A_0}}-\bm{D}\bm{\tilde{p}}-\bm{\hat{Z}} 
t^{5/2}  \bm{{\rm Li}_{\frac{5}{2}}}.
\label{eq:result_A}
\end{align}
The last term in the function $\bm{\tilde{A}}$ is negligible in pressure. Nevertheless, it is 
necessary in the  calculation of the scaling function or phase boundaries. 
After a lengthy iteration, we get a close form  of the EOS 
 \begin{align}
\tilde{\bm{p}}
=&-\bm{M_r} \bm{F_{\frac{3}{2}}},
\label{eq:result_sun_p}\\
\bm{\tilde{A}}
\approx & \bm{\tilde{A_0}}+\bm{D} \bm{M_r} \bm{f_{\frac{3}{2}}}
+\bm{D} \bm{M_r} \bm{f_{\frac{1}{2}}} \bm{D} \bm{M_r} \bm{f_{\frac{3}{2}}}.\nonumber
\end{align}

Furthermore,  we could obtain all the 
thermodynamic quantities of the system in equilibrium by standard thermodynamic relations via  the  pressure  Eq.(\ref{eq:result_p}), which  serves as  the grand thermodynamic potential of
the system. In this context,  the partial derivatives of the pressure by any chemical potential, external  fields and temperature are essential in our approach. Thus we take the 
derivative of the pressure Eq.(\ref{eq:result_sun_p})  with respect to the variable $\eta$:
$\eta=\tilde{\mu},h_1,h_2,...$.  It follows that 
\begin{align}
\frac{\partial {\tilde{\mathbf{p}}}}{\partial \eta}
=&-\bm{M_r}\mathbf{F_{\frac{1}{2}}} {\frac{\partial \tilde{\mathbf{A}}}{\partial \eta}},\\
{\frac{\partial \tilde{\mathbf{A}}}{\partial \eta}}
=& {\frac{\partial \tilde{\mathbf{A}}_0}{\partial \eta}}-\mathbf{D}
\frac{\partial \tilde{\mathbf{p}}}{\partial \eta}
-\bm{\hat{Z}} 
\mathbf{F_{\frac{3}{2}}} {\frac{\partial \tilde{\mathbf{A}}}{\partial \eta}}.
\end{align}
By solving the above two linear equations,  we obtain the first order derivative  thermodynamic properties
{\small 
\begin{align}
\frac{\partial \tilde{\mathbf{p}}}{\partial \eta}
=&-\bm{M_r F_{\frac{1}{2}}}(\mathbf{I}+\bm{D}\bm{M_r F_{\frac{1}{2}}}
+(\bm{D}\bm{M_r F_{\frac{1}{2}}})^2)
\frac{\partial \tilde{\mathbf{A}}_0}{\partial \eta}, 
\label{eq:result_pde} \\
\frac{\partial \bm{\tilde{A}}}{\partial \eta}
=&(\mathbf{I}+\bm{D}\bm{M_r F_{\frac{1}{2}}}
+(\bm{D}\bm{M_r F_{\frac{1}{2}}})^2)
\frac{\partial \tilde{\mathbf{A}}_0}{\partial \eta}.\nonumber 
\end{align}
}

For the second order phase transitions, the second derivatives of the pressure, 
the compressibility or the susceptibility for instance, give a deep insight into the quantum criticality of the systems. Similarly,  the second order thermodynamic quantities can  be obtained 
\begin{eqnarray}
\frac{\partial^2 \bm{\tilde{p}}}{\partial \eta_1 \partial \eta_2}
&=&\left[-\bm{M_r}\bm{F_{\frac{1}{2}}}(\bm{I}+\bm{D}\bm{M_r}\bm{F_{\frac{1}{2}}})
\bm{D}\bm{M_r}\bm{F_{-\frac{1}{2}}}
\right.\nonumber\\
&&
\left.-\bm{M_r}\bm{F_{-\frac{1}{2}}} \right]
\left(\frac{\partial \bm{\tilde{A}}}{\partial \eta_1} 
\frac{\partial \bm{\tilde{A}}}{\partial \eta_2}\right).
\label{eq:result_pde2}
\end{eqnarray}

On the other hand, the derivative of pressure with respect to $t$ always imposes a tedious task. 
After   carefully solving the above equations involving the derivatives of pressure,  we obtain the entropy $\tilde{\bm{s}}=\frac{\partial \bm{\tilde{p}}}{\partial t}$
\begin{eqnarray}
\frac{\partial \bm{\tilde{p}}}{\partial t}
&=&-\bm{M_r}\frac{3}{2t}\bm{F_{\frac{3}{2}}}
+\bm{M_r}\bm{F_{\frac{1}{2}}}\frac{\bm{\tilde{A}}}{t} -\bm{M_r}\bm{F_{\frac{1}{2}}}(\bm{I}+\bm{D M_r F_{\frac{1}{2}}})
\notag \\
&&
\times (-\bm{D M_r F_{\frac{1}{2}}}\frac{\bm{\tilde{A}}}{t}
+\bm{D M_r} \frac{3}{2t}\bm{F_{\frac{3}{2}}}),
\label{eq:result_pdt} \\
\frac{\partial \bm{\tilde{A}}}{\partial t}
&=&\left[\mathbf{I}+\bm{D}\bm{M_r F_{\frac{1}{2}}}
+(\bm{D}\bm{M_r F_{\frac{1}{2}}})^2\right] \left(-\bm{D M_r F_{\frac{1}{2}}}\frac{\bm{\tilde{A}}}{t}\right.
\notag \\
&&
\left. 
+\bm{D M_r} \frac{3}{2t}\bm{F_{\frac{3}{2}}}
+\bm{\hat{Z}F_{\frac{3}{2}}}\frac{\bm{\tilde{A}}}{t}
-\bm{\hat{Z}}\frac{5}{2t}\bm{F_{\frac{5}{2}}}\right).
\end{eqnarray}
This result contains not only the linear-temperature-dependent  behaviour of the entropy  in the Luttigner liquid (for $T\ll E_F$) but also the universal quantum scalings  of the entropy in the quantum critical region (for $T\gg E_F$) in the vicinity of the critical point. 
Similarly, the second derivative of the pressure with respect to $t$ is given by 
\begin{eqnarray}
\frac{\partial^2 \bm{\tilde{p}}}{\partial t^2}
&=&-\bm{M_r}\left( \frac{3}{4t^2} \bm{F_{\frac{3}{2}}} 
+\frac{1}{t} \bm{F_{\frac{1}{2}}}\bm{B}+\bm{F_{-\frac{1}{2}}}\bm{B^2}+\right.\nonumber\\
&&
\left. \bm{F_{\frac{1}{2}}}\bm{D M_r}(\frac{3}{4t^2} \bm{F_{\frac{3}{2}}} 
+\frac{1}{t} \bm{F_{\frac{1}{2}}}\bm{B}+\bm{F_{-\frac{1}{2}}}\bm{B^2})
\right)
\label{eq:result_pdt2}
\end{eqnarray}
with $\bm{B}=\frac{\partial \bm{\tilde{A}}}{\partial t}-\frac{\bm{\tilde{A}}}{t}$. 
In contrast to the previous studies on multicomponent  interacting fermions \cite{GuaBL13,Yu:2016,WuHZ03,CazR14,Schlottmann:1993,LeeGBY11,Jiang:2016,He:2010,Oelkers:2006}, the above close forms of the thermodynamics are very useful for analyzing the behaviour of the  quantum liquids and critical scalings of the 1D interacting fermions with  $SU(w)$ symmetry. 

{\bf Quantum criticality.}
In the vicinities of  the phase boundaries in the phase diagrams of the the 1D interacting fermions with  $SU(w)$ symmetries \cite{Yu:2016},  a discontinuity emerges in the polylog functions ${\rm{Li}_s(x)}$ in the EOS, namely,
\begin{equation}
\lim_{\mu/t \to 0^+} {\rm Li}_s (-e^{\mu/t})=-\frac{(\mu/t)^s}{{\rm \Gamma}(s+1)},\,\,
\lim_{\mu/t \to 0^-} {\rm Li}_s (-e^{\mu/t})=0.
\end{equation}
The sign change of the $\tilde{A}^{(r)}$ leads to a sudden change of the Polylog functions at the 
critical point of a phase transition, for example, see Fig.~\ref{fig:Phase_diagram}, where a sudden enhancement  in the Wilson ratio is observed when the driving parameters is tuned across any  phase boundary. 
The condition $A^{(r)}>0$ implies the  exist of the bound states of $r$-fermions in this certain  quantum phase. 
As a consequence, in the vicinities of  phase boundaries, any  thermodynamic quantity can be separated into 
 two parts: 
\begin{enumerate}
\item the background part and 
\item the discontinuous part. 
\end{enumerate}
The background part involves the states which do not have a sudden change and
the calculation of this part is cumbersome \cite{Guan-Ho}.
The discontinuous part can be obtained by analyzing  the first order 
of divergence in the EOS, i.e. the part involves a sudden change in the density of  a certain branch of bound states. For example, at the phase transition from the fully paired states into the FFLO phase, the regular part mainly relates to  the thermal fluctuation presented by functions with $A^{(2)}$, whereas the singular part results in a sign change of  $A^{(1)}$ ,which indicates the crossover of the density of unpaired fermions from zero to non-zero 
while the Fermi sea of the unpaired fermions starts to fill with particles. 

\begin{figure}[t]
\begin{center}
  \includegraphics [width=1.0\linewidth]{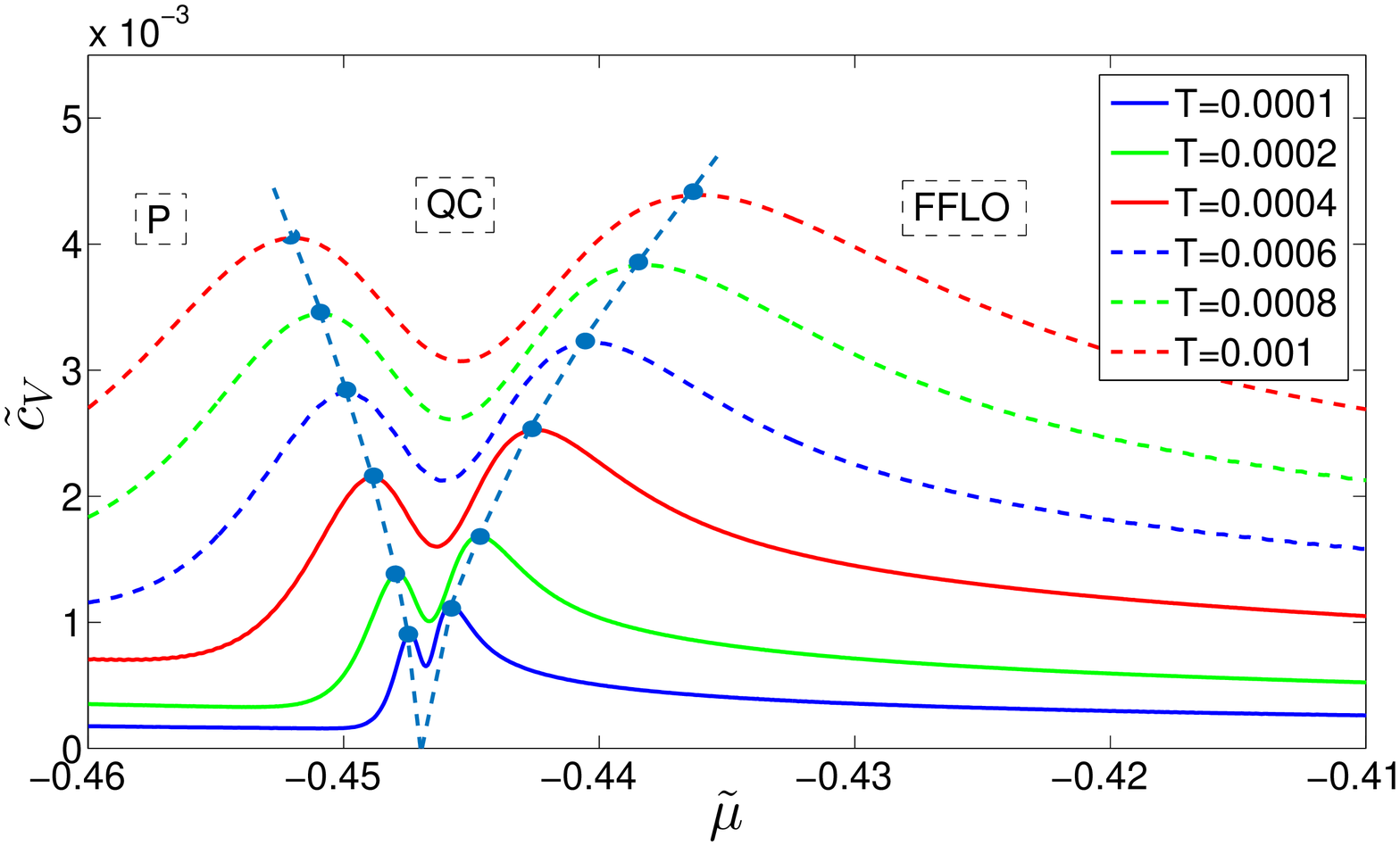}
\end{center}
 \caption{The temperature rescaled specific heat $c_v/T $ is in the unit $\frac{1}{2}|c|$ vs the dimensionless chemical potential $\tilde{\mu}=\frac{\mu}{\frac{1}{2}|c|^2}$ for a fixed dimensionless external field $ \tilde{h}=0.46$ in the $r=2$ Fermi gas (\ref{Ham}). The phase transition occurs when  the chemical potential drives the system across the phase boundary from fully-paired phase  into the FFLO-like state at $t=0$. The specific peaks fanning out from the critical point  indicate three regions: the Luttigner liquid of bosonic bound pairs (P), quantum critical (QC) region  and the phase of the FFLO like state.
  }
 \label{fig:cv}
\end{figure}

We further calculate the phase transition driven by chemical potential or external fields. Supposing  that  the quantum phase transition is driven by
the sign change of  $\tilde{A}^{(r)}$ ,  
we observe that the thermodynamic quantities are naturally  split into background and discontinuous parts as:
\begin{align}
\frac{\partial \tilde{\mathbf{p}}}{\partial \eta}
=&-\bm{M_r F_{\frac{1}{2}}}(\mathbf{I}+\bm{D}\bm{M_r F_{\frac{1}{2}}}
+(\bm{D}\bm{M_r F_{\frac{1}{2}}})^2)
\frac{\partial \tilde{\mathbf{A}}_0}{\partial \eta}  \notag \\
=&\frac{\partial \tilde{\mathbf{p}}}{\partial \eta} \Big|_0
-\bm{M_r F_{\frac{1}{2}}}\frac{\partial \tilde{\mathbf{A}}_0}{\partial \eta},\label{CR1}
\end{align}
where $\frac{\partial \tilde{\mathbf{p}}}{\partial \eta} \Big|_0$ denotes  the backgroud part, i.e. 
{\small 
\begin{align}
\frac{\partial \tilde{\mathbf{p}}}{\partial \eta} \Big|_0
=-\bm{M_r F_{\frac{1}{2}}}\left(\mathbf{I}+\bm{D}\bm{M_r F_{\frac{1}{2}}}
+(\bm{D}\bm{M_r F_{\frac{1}{2}}})^2\right)
\frac{\partial \tilde{\mathbf{A}}_0}{\partial \eta} 
\Big|_{F_s^{(r)}=0}.\nonumber
\end{align}
}
Then we obtain the explicit scaling form of the first derivative thermodynamic quantities of the $SU(w)$  Fermi gases 
\begin{align}
\frac{\partial \tilde{p}}{\partial \eta} 
=&\frac{\partial \tilde{p}}{\partial \eta} \Big|_0^{(r)}
-\frac{\sqrt{r}}{2\sqrt{\pi}} \frac{\partial \tilde{A}^{(r)}_0}{\partial \eta} 
t^{1/2} {\rm Li}_\frac{1}{2}\left(-e^{r(\tilde{\mu}-\mu_{cr})/t}\right),\label{CR2}
\end{align}
where the $\mu_{cr}$ is the critical chemical potential for the phase transition induced by the change of  the 
$r$-atoms bond states.  This method can  be further applied to  the second order thermodynamic quantities, 
explicitly 
\begin{eqnarray}
\frac{\partial^2 {\tilde{p}}}{\partial \eta_1 \partial \eta_2}\label{CR3}
&\approx &\frac{\partial^2 {\tilde{p}}}{\partial \eta_1 \partial \eta_2} \Big|_0^{(r)}\\
&&
-\frac{\sqrt{r}}{2\sqrt{\pi}} \frac{\partial \tilde{A}^{(r)}_0}{\partial \eta_1} 
\frac{\partial \tilde{A}^{(r)}_0}{\partial \eta_2}
t^{-1/2} {\rm Li}_{-\frac{1}{2}}(-e^{r(\tilde{\mu}-\mu_{cr})/t}).\nonumber
\end{eqnarray}
We thus read off the critical exponents from these scaling functions, i.e. the dynamic critical exponent $z=2$ and the correlation critical exponent $\nu=1/2$. 
In particular,  the specific heat is given by 
\begin{eqnarray}
\frac{\tilde{c}_V}{t}&\approx& \frac{\partial^2 \bm{\tilde{p}}}{\partial t^2}
\approx \frac{\partial^2 \bm{\tilde{p}}}{\partial t^2} \Big|_0\nonumber \\
&&
-\bm{M_r}\left( \frac{3}{4t^2} \bm{F_{\frac{3}{2}}} 
-\frac{1}{t} \bm{F_{\frac{1}{2}}}{\frac{\tilde{\bm{A}}_0}{t}}
+\bm{F_{-\frac{1}{2}}}\frac{\tilde{(\bm{A}}_0)^2}{t^2}
\right)
\end{eqnarray}
that gives 
\begin{align}
\frac{\partial^2 \tilde{p}}{\partial t^2}
\approx  &\frac{\partial^2 \tilde{p}}{\partial t^2} \Big|_0^{(r)} 
-\frac{\sqrt{r}}{2\sqrt{\pi t}} \mathcal{H} \left(\frac{r(\tilde{\mu}-\mu_{cr})}{t}\right),\label{CR4}
\end{align}
where the function 
\begin{equation}
\mathcal{H} (x) = \frac{3}{4} {\rm Li}_{\frac{3}{2}} ( - e^x) - x
{\rm Li}_{\frac{1}{2}} ( - e^x) + x^2 {\rm Li}_{- \frac{1}{2}} ( - e^x).
\end{equation}
Solving  the equation $\frac{d}{dx} \mathcal{H} (x)=0$, we get two solutions:
$y_1\approx 0.639844,\,y_2 \approx 0.276201$ that  determine  the two peaks of the 
specific heat at quantum criticality in the $SU(w)$ Fermi gases, i.e. 
\begin{equation}
t^*_1=-y_1\,r(\tilde{\mu}-\mu_{cr}),\quad t^*_1=y_2\, r(\tilde{\mu}-\mu_{cr}).
\end{equation}
The two crossover temperatures fanning out from the critical point indicate  the quantum critical region beyond  the quantum liquid phases, see Fig.~\ref{fig:cv}.
Recent studies on the quantum criticality in 1D  Heisenberg spin chain \cite{He:2017} and 1D Bose gas \cite{Exp-B6} confirm such a novel  existence of the critical cone while a quantum phase transition occurs.

For clarity and possible experimental use, we present the explicit  scaling forms of the thermodynamic properties for the 1D strongly attractive $SU(w)$ Fermi gases. The polarization of the systems is $ \tilde{m}=\sum_k^w \frac{1}{2} k(w-k) \tilde{n}_k$, 
where $\tilde{n}_r=\frac{\partial \tilde{p}}{\partial h_r}$ presents the density of the 
$r$-atoms . 
For the phase transition related to the sign change of  $\tilde{A}^{(r)}$, i.e. the phase transition occurs when  the Fermi sea of the charge bound states of $r$-atoms starts to fill or to be gapped  at zero temperature. Based on the above unified scaling forms (\ref{CR1})--(\ref{CR4}), 
we can directly present  the result of the scaling functions of  physical quantities
\begin{align}
\tilde{n} \approx & \tilde{n}_0 -\frac{r\sqrt{r}}{2\sqrt{\pi}} t^{1/2}
 \mathcal{F} (\frac{r(\tilde{\mu}-\mu_{cr})}{t})  \notag \\
\tilde{m} \approx & \tilde{m}_0 -\frac{1}{2} (w-r)r
\frac{\sqrt{r}}{2\sqrt{\pi}} t^{1/2}
 \mathcal{F} (\frac{r(\tilde{\mu}-\mu_{cr})}{t}) \notag \\
\tilde{\kappa} \approx & \tilde{\kappa}_0 
-\frac{r^2\sqrt{r}}{2\sqrt{\pi}} t^{-1/2}
 \mathcal{G} (\frac{r(\tilde{\mu}-\mu_{cr})}{t}) \notag  \\
 \tilde{\chi} \approx & \tilde{\chi}_0
 -\frac{1}{2} r(w-r) \frac{\sqrt{r}}{2\sqrt{\pi}} t^{-1/2}
 \mathcal{G} (\frac{r(\tilde{\mu}-\mu_{cr})}{t}) \notag  \\
 \frac{\tilde{c}_V}{t} \approx &\frac{\tilde{c}_{V0}}{t}
 -\frac{\sqrt{r}}{2\sqrt{\pi }} t^{-1/2} 
 \mathcal{H} \left(\frac{r(\tilde{\mu}-\mu_{cr})}{t}\right) 
 \notag 
\end{align}
with 
\begin{align}
\mathcal{F}(x) =& {\rm Li}_{\frac{1}{2}} ( - e^x) \notag \\
\mathcal{G}(x) =& {\rm Li}_{-\frac{1}{2}} ( - e^x) \notag \\
\mathcal{H} (x) = &\frac{3}{4} {\rm Li}_{\frac{3}{2}} ( - e^x) - x
{\rm Li}_{\frac{1}{2}} ( - e^x) + x^2 {\rm Li}_{- \frac{1}{2}} ( - e^x). \notag
\end{align}
Here every first term in the above quantities denote the background parts. 
Notice that for the  $SU(2)$ case, we conventionally  take the magnetic field $H=2H_1$, and the dimensionless form now $
\tilde{p}^{(r)}=\frac{p^{(r)}}{\frac{1}{2}|c|^3}, 
\tilde{A}^{(r)}=\frac{A^{(r)}}{\frac{1}{2}|c|^2},
\tilde{\mu}=\frac{\mu}{\frac{1}{2}|c|^2},
{h}=\frac{H}{\frac{1}{2}|c|^2},
t=\frac{T}{\frac{1}{2}|c|^2}
$.
These scaling functions provide exact result of  quantum critical phenomena of  the 1D $SU(w)$ Fermi gases, also see the theory of quantum criticality   \cite{QC-Book,Giamarchi:2004}.

In summary, we have presented a unified approach to the thermodynamics and quantum scaling functions for the 1D strongly attractive Fermi gases with $SU(w)$ symmetry. In particular, we have obtained the two crossover temperature lines fanning out from the critical point that confirm the existence of the critical cone at quantum criticality. While the quantum liquids can be measured through the dimensionless ratios, revealing the  important free fermion nature of 1D interacting fermions. Our results pave a way to experimentally study quantum criticality of 
the fermionic alkaline-earth atoms that display an exact $SU(w)$ spin symmetries with $w=2I+1$ \cite{Gor2010,Cazalilla:2009}. Here $I$ is the nuclear spin. The study of critical phenomena and quantum correlations in ultracold atoms with  high symmetries has become a new frontier in atomic physics. 

\noindent

{\em Acknowledgments.}  The authors  thank Yu-Peng Wang,  Yu-Zhu Jiang, Feng He, Song Cheng   and Peng He   for helpful discussions. This work is supported by the NSFC under grant numbers 11374331 and the key NSFC grant No.\ 11534014. XWG has been partially supported by the Australian Research Council.

\end{document}